\documentclass[3p,times]{elsarticle}

\usepackage{ecrc}

\usepackage{epstopdf}

\volume{00}

\firstpage{1}

\journalname{Nuclear Physics A}

\runauth{L. Ruan}

\jid{nupha}

\jnltitlelogo{Nuclear Physics A}

\usepackage{graphicx}
\usepackage{amsmath,amssymb}
\begin{document}
%\begin{document}
\def\Journal#1#2#3#4{{#1} {\bf #2}, #3 (#4)}

% Some useful journal names
\def\NCA{Nuovo Cimento}
\def\NIM{Nucl. Instr. Meth.}
\def\NIMA{{Nucl. Instr. Meth.} A}
\def\NPB{{Nucl. Phys.} B}
\def\NPA{{Nucl. Phys.} A}
\def\PLB{{Phys. Lett.}  B}
\def\PRL{Phys. Rev. Lett.}
\def\PRC{{Phys. Rev.} C}
\def\PRD{{Phys. Rev.} D}
\def\ZPC{{Z. Phys.} C}
\def\JPG{{J. Phys.} G}
\def\CPC{Comput. Phys. Commun.}
\def\EPJ{{Eur. Phys. J.} C}
\def\PR{Phys. Rept.}
\def\JHEP{JHEP}

\begin{frontmatter}



\title{The low and intermediate mass dilepton and photon results}

%% Single author (and collaboration) - please insert
\author{Lijuan Ruan}

\address{Physics Department, Brookhaven National laboratory, Upton NY 11973}

%% For multiple authors, replace the above by:

%\author[label1]{Author1}
%\author[label2]{Author2}

%\address[label1]{Address 1}
%\address[label2]{Address 2}

\begin{abstract}
   I summarize and discuss some of the experimental results on the low and intermediate mass dileptons and direct photons presented at {\it Quark Matter 2014}.

\end{abstract}


\end{frontmatter}

% \linenumbers


\section{Introduction}
\label{intro}
Ultra-relativistic heavy ion collisions provide a unique
environment to study the properties of strongly interacting matter
at high temperature and high energy density~\cite{starwhitepaper}.
Leptons and photons are penetrating probes of the hot, dense medium since they are not affected by the strong interaction and therefore they can probe the whole evolution of the
collision.

In the low invariant mass range of produced lepton pairs
($M_{ll}\!<\!1.1$ GeV/$c^{2}$), we can study vector meson
in-medium properties through their dilepton decays, where
modifications of mass and width of the spectral functions observed
may relate to the possibility of chiral symmetry
restoration~\cite{dilepton,dileptonII}. The dilepton spectra in the intermediate mass range
($1.1\!<M_{ll}\!<\!3.0$ GeV/$c^{2}$) are directly related to
thermal radiation of the Quark-Gluon Plasma
(QGP)~\cite{dilepton,dileptonII}. However, contributions from
other sources have to be obtained experimentally. Such
contributions include background pairs from correlated open heavy
flavor decays 
($c\bar{c}\rightarrow l^{+}l^{-}X$ or $b\bar{b}\rightarrow
l^{+}l^{-}X$). In addition, photons in the low transverse momentum range $1\!<p_{T}\!<\!4$ GeV/$c$  are used to study thermal radiation from QGP and hadronic gas.

In this article, I will summarize the results from NA60, PHENIX, STAR, and ALICE on thermal dileptons and photons presented at {\it Quark Matter 2014}.

\section{Dileptons}
\subsection{Results on $p+p$ and $p(d)+A$ collisions}
At this conference, the STAR Collaboration presented their 200 GeV $p+p$ reference measurement using high statistics 2012 data~\cite{Yang:14}. The cocktail simulation with expected hadronic contribution is consistent with the data. A similar conclusion holds for $p+p$ collisions at LHC energies~\cite{Kohler:14}. In addition, in d+Au and p+Pb collisions, hadronic cocktails are also found to be consistent with the measurements, indicating that there is no medium radiation observed there~\cite{Kohler:14,Dion:14}.

\subsection{Thermal dileptons in the low mass region}
At the SPS, the low mass
dilepton enhancements in the CERES $e^+e^-$ data ~\cite{ceres} and
in the NA60 $\mu^+\mu^-$ data~\cite{na60,na60:10} indicate substantial
medium effects on the $\rho$-meson spectral function. The precise
NA60 measurement of the low mass enhancement provides a decisive
discrimination between the dropping-mass scenario~\cite{dropmass}
and the massively broadened spectral function~\cite{massbroaden}.
The latter one was found to be able to describe the
data consistently. 

At RHIC, the PHENIX experiment observed a significant enhancement
in the $e^{+}e^{-}$ data above the expectation from hadronic
sources for $0.15\!<\!M_{ee}\!<\!0.75$ GeV/$c^{2}$ and $p_T\!<1$
GeV/$c$ in 200 GeV Au+Au collisions~\cite{lowmass}. Models~\cite{rapp:09,PSHD:12,USTC:12} that successfully
describe the SPS dilepton data, fail to describe the
PHENIX data. After the Time-of-Flight detector upgrade, STAR
reported the dielectron spectra in 200 GeV Au+Au collisions at
QM2011 and the low mass excess was not as significant as what
PHENIX observed in 0-80\% and 0-10\% collisions~\cite{ppdilepton:12,Jie:11}. Further
comparisons point to the fact that the discrepancy between STAR
and PHENIX comes from 0-20\% central collisions only~\cite{GaleRuan:12}.

\begin{figure}[htbp]
\begin{center}
\includegraphics[width=0.80\textwidth]{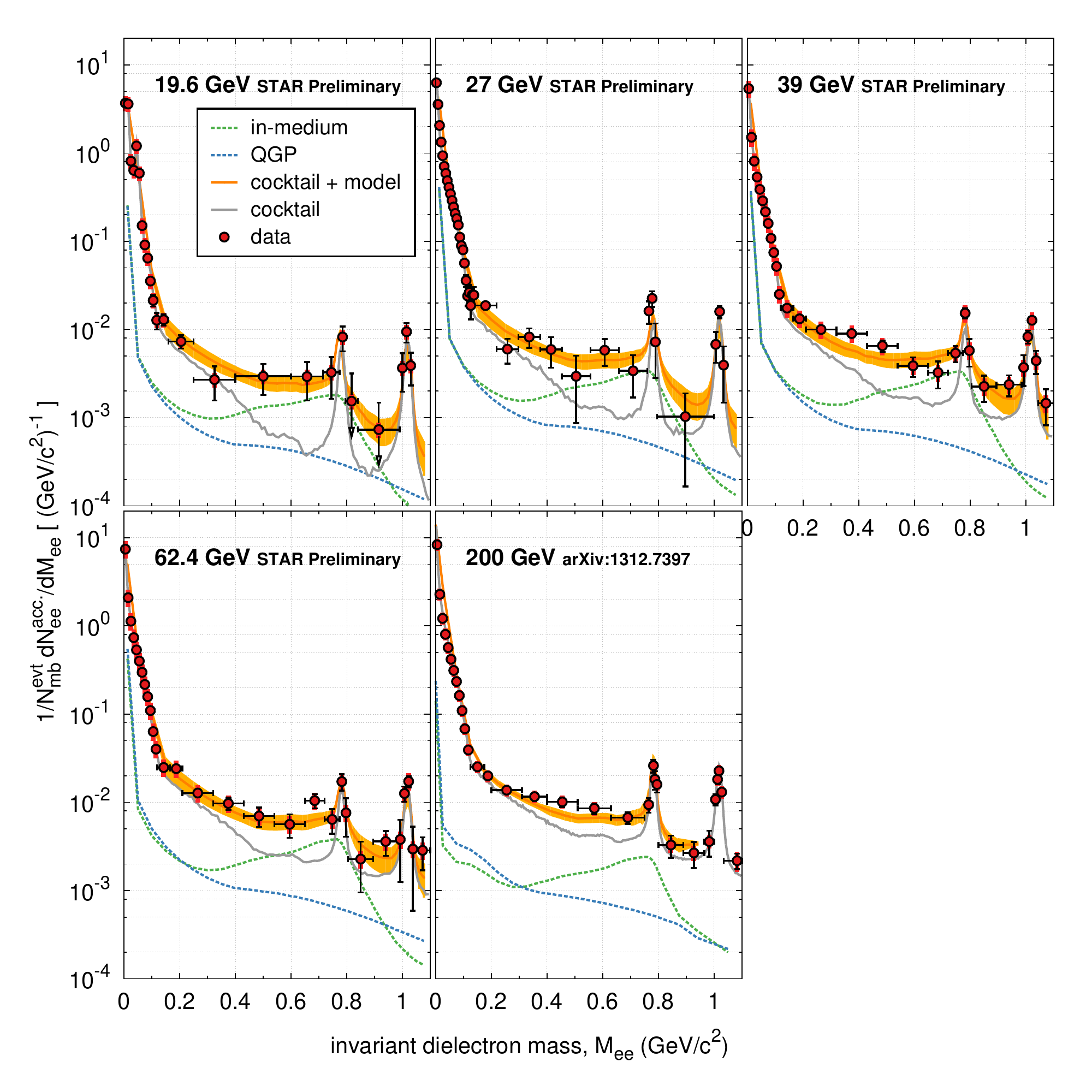}
\end{center}
\caption{(Color Online) Dielectron spectra in Au+Au collisions
from 19.6, 27, 39, 62.4, and 200 GeV from STAR~\cite{Huck:14}. In each panel, the "cocktail" represents the expected hadronic contribution~\cite{Huck:14,stardielectron:14} while "in-medium" and "QGP" represent contributions from a broadened $\rho$ spectral function and QGP thermal radiation, respectively~\cite{rapp:09}. The sum of the "cocktail", "in-medium", and "QGP" is shown as the solid curve on top of the data points.}
\label{fig:1}
\end{figure}

At Quark Matter 2012, PHENIX reported the dielectron results from
the Hadron Blind Detector from 20-92\% Au+Au collisions at 200
GeV, which are consistent with the data in the corresponding centralities 
in the previous publication~\cite{phenixdielectron:12}.
STAR reported the dielectron spectra from 19.6, 39, and 62.4 GeV
Au+Au collisions~\cite{geurts:12,huang:12}.  See also Ref. \cite{GaleRuan:12}.

At this conference, STAR reported the dielectron spectrum from 27 GeV Au+Au collisions~\cite{Huck:14}.  The 200 GeV Au+Au results were finalized~\cite{stardielectron:14}. In addition, the $p_T$ differential measurements were also shown in Au+Au collisions at 19.6, 27, 39, and 62.4 GeV~\cite{Huck:14}. 
A broadened spectral function~\cite{rapp:09}, which describes
SPS dilepton data, consistently accounts for the STAR low mass
excess in Au+Au collisions at 19.6, 27, 39, 62.4, and 200 GeV, as shown in Fig.~\ref{fig:1}.  It also describes the $p_T$ dependence of low mass excess up to 2 GeV/$c$ for all these collision energies~\cite{Huck:14}. In addition, the excess dielectron mass spectrum in the mass region 0.3-0.76 GeV/$c^2$ in 
200 GeV Au+Au collisions is found to follow $N_{part}^{1.54\pm0.18}$ dependence, where $N_{part}$ is the number of participant nucleons in a collision~\cite{stardielectron:14}.

\subsection{Thermal dileptons in the intermediate mass region}
\begin{figure}[htbp]
\begin{center}
\includegraphics[width=0.7\textwidth]{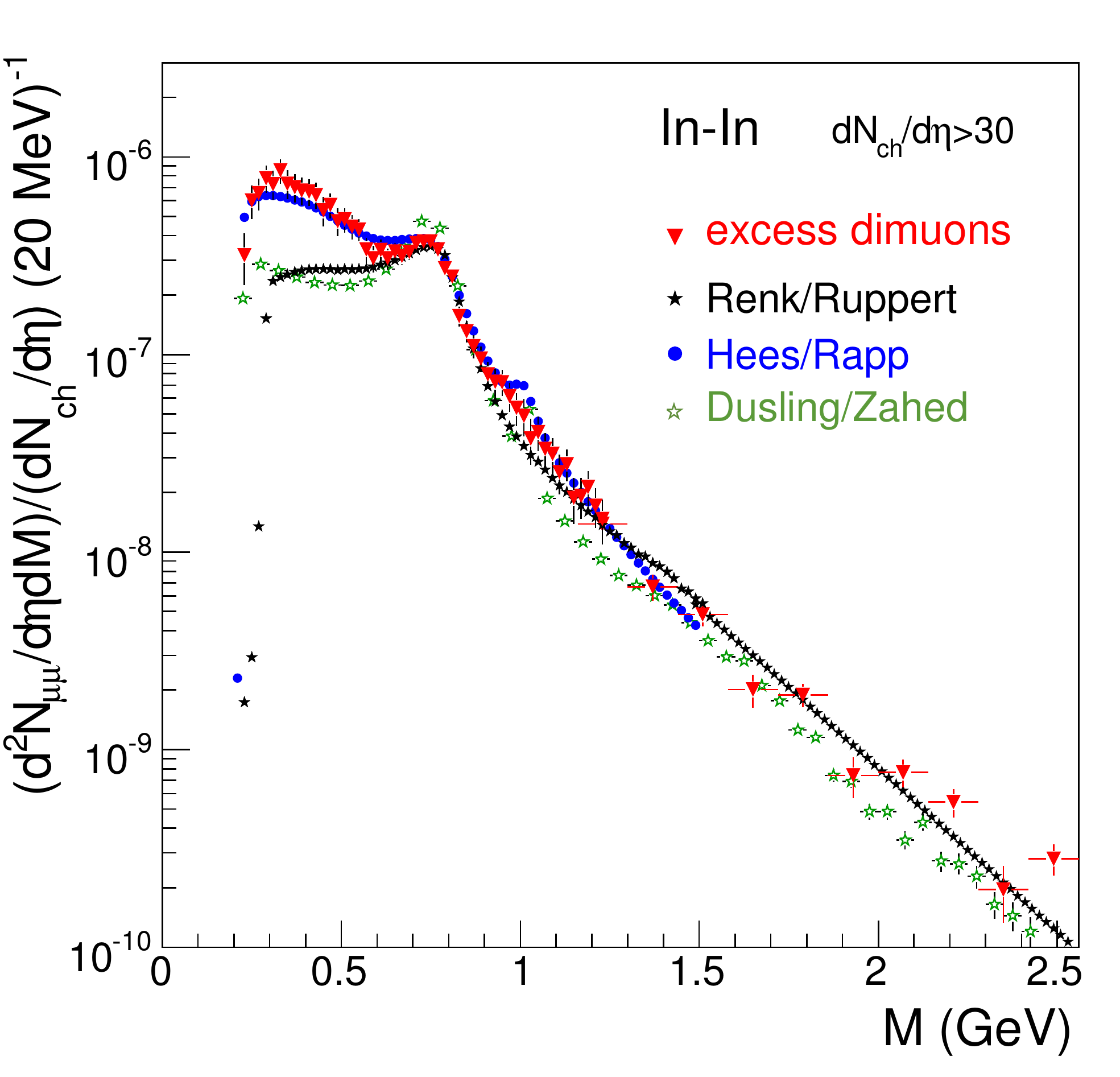}
\end{center}
\caption{(Color Online) Acceptance-corrected excess dimuon invariant mass spectrum in In+In collisions at 17.3 GeV. Model comparisons are also shown~\cite{na60:10}. }
\label{fig:2}
\end{figure}

Dilepton spectra in the intermediate mass range
are directly related to the thermal radiation of the QGP~\cite{dilepton,dileptonII}. Thanks to the fact, that the charm cross section at 17.3 GeV is small and with a vertex detector to reject the  charm background, NA60 presented the unprecedented excess dimuon mass spectrum in In+In collisions, as shown in Fig.~\ref{fig:2}. From its intermediate mass region, one can obtain the temperature of the emitting source. In addition, more differential results such as the $p_T$, azimuthal angle, and polar angle dependences of the excess dimuons in different mass regions were obtained~\cite{na60:10}. Based on those precise measurements, it was identified that the excess dimuons in the intermediate mass region comes from the early emission from the hot, dense medium.

At RHIC, with the data sets currently available, it is difficult to measure correlated charm
contribution or QGP thermal radiation in the
intermediate mass region. At RHIC energies, there is so far 
no clear statement about thermal radiation in the
intermediate mass region. The recent STAR measurements in 200 GeV Au+Au indicate that the trend there could be slightly different in 0-80\% and 0-10\% collisions~\cite{stardielectron:14}. However, the results are not precise enough to draw a conclusion.
The recent detector upgrade with the
Heavy Flavor Tracker at STAR, completed in 2014, will
provide precise charm cross section measurements~\cite{hft}. This
will help to understand heavy quark dynamics in the medium and
constrain model inputs to calculate dilepton mass spectra from correlated charm
contribution. However the measurements of $c\bar{c}$ correlations
will still be challenging if not impossible. An independent
approach is planned with the Muon Telescope Detector upgrade
(MTD)~\cite{starmtdproposal}, which was completed in 2014. The $\mu-e$ correlations will measure the contribution from heavy flavor correlations to the dielectron or
dimuon continuum with the 200 GeV Au+Au data taken in 2014~\cite{starmtdproposal,Ruan:12}. This will allow to access the thermal radiation in the intermediate mass region.

In addition, the Beam Energy Scan Phase II at RHIC in 2018-2019 will enable to collect sufficient data in Au+Au collisions from 7.7 to 19.6 GeV~\cite{besII:14}. The charm cross section is significantly reduced at lower energies.  One would like to see whether at some collision energies the transition from low to intermediate dilepton mass region would be smooth, which would imply chiral symmetry restoration. 

\section{Thermal photon measurements}
\subsection{Review of previous results}
Photons at $1\!<\!p_{T}\!<\!4$ GeV/$c$ are used to study thermal radiation from QGP and hadronic medium. For
$1\!<\!p_{T}\!<\!4$ GeV/$c$, PHENIX measured direct photon yields
from dielectron measurements and found an excess in 0-20\% Au+Au over $p+p$ at $\sqrt{s_{_{NN}}} = 200$ GeV, exponential in $p_T$ with the inverse slope parameter
221 MeV~\cite{thermalphoton}. At LHC, an excess of direct photon yield in 0-40\% Pb+Pb
collisions above $p+p$ was reported at $\sqrt{s_{_{NN}}} = $ 2.76
TeV, exponential in $p_T$ with the inverse slope parameter 304 MeV~\cite{ALICEphoton}. If
indeed the excess is from the QGP phase, the measurements at RHIC and
LHC would indicate that the initial temperature of the QGP
evolution is as high as 300-600 MeV~\cite{thermalphoton}.

On the other hand, $v_2$ of direct photons has been 
found to be substantial in the range $1\!<\!p_{T}\!<\!4$ GeV/$c$
in central 0-20\% Au+Au collisions at $\sqrt{s_{_{NN}}} = 200$
GeV~\cite{photonv2}. Model calculations~\cite{Rupa:09} for QGP
thermal photons in this kinematic region significantly
under-predict the observed $v_2$, while if a significant
contribution from the hadronic sources at later stages is added,
the excess of the spectra and the observed $v_2$ at
$1\!<\!p_{T}\!<\!4$ GeV/$c$ are described reasonably
well~\cite{rapp:11}. In addition, ALICE reported that significant $v_2$ is observed for direct photons, though the systematic uncertainties are large~\cite{ALICEv2:12}.

\subsection{Review of new results}

\begin{figure}[htbp]
\begin{center}
\includegraphics[width=0.7\textwidth]{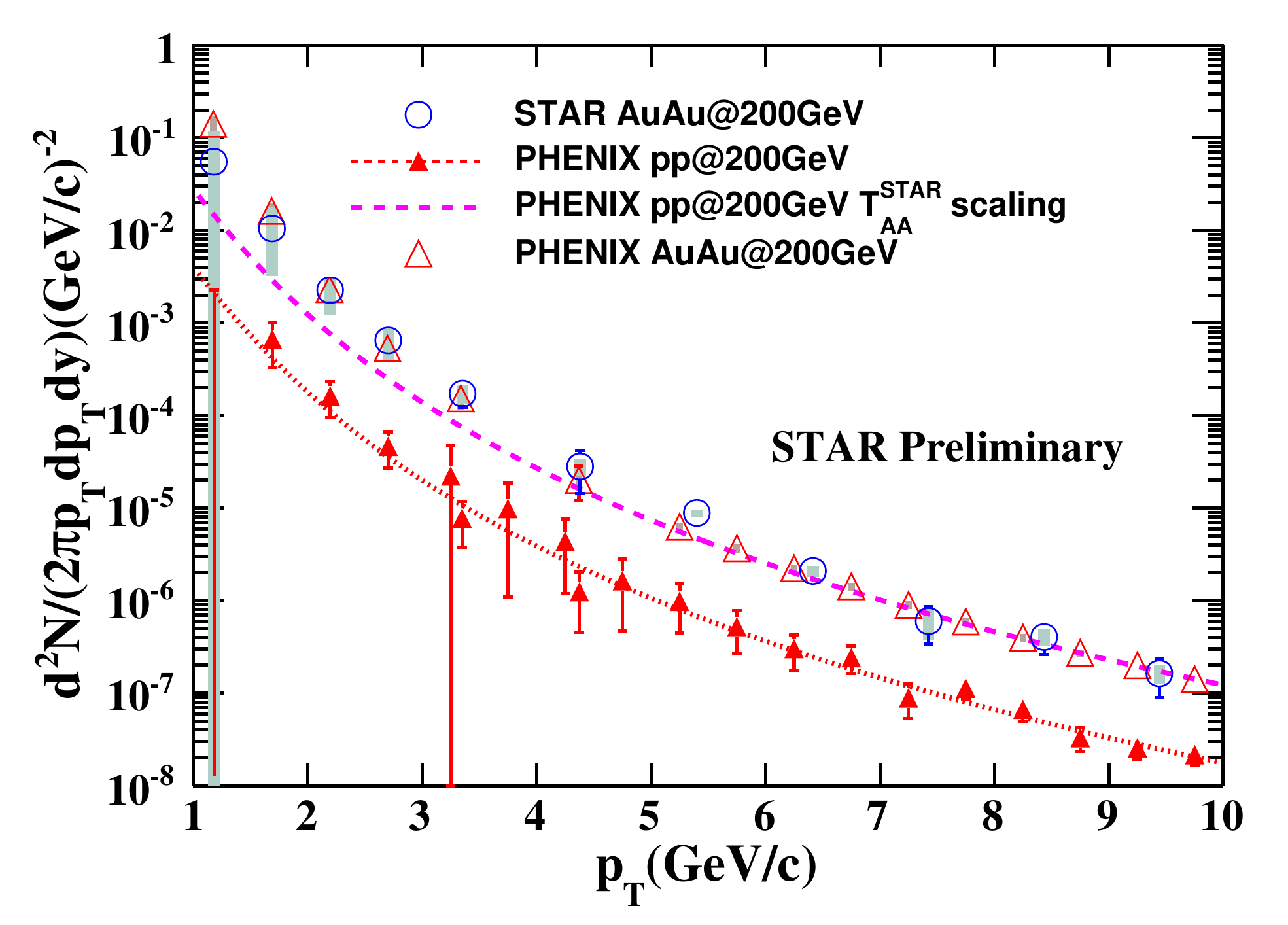}
\end{center}
\caption{(Color Online) Direct photon yields as a function of $p_T$ in 0-80\% Au+Au  from STAR, 0-92\% Au+Au from PHENIX, and $p+p$ from PHENIX at 200 GeV~\cite{Yang:14}. }
\label{fig:3}
\end{figure}

\begin{figure}[htbp]
\begin{center}
\includegraphics[width=0.95\textwidth]{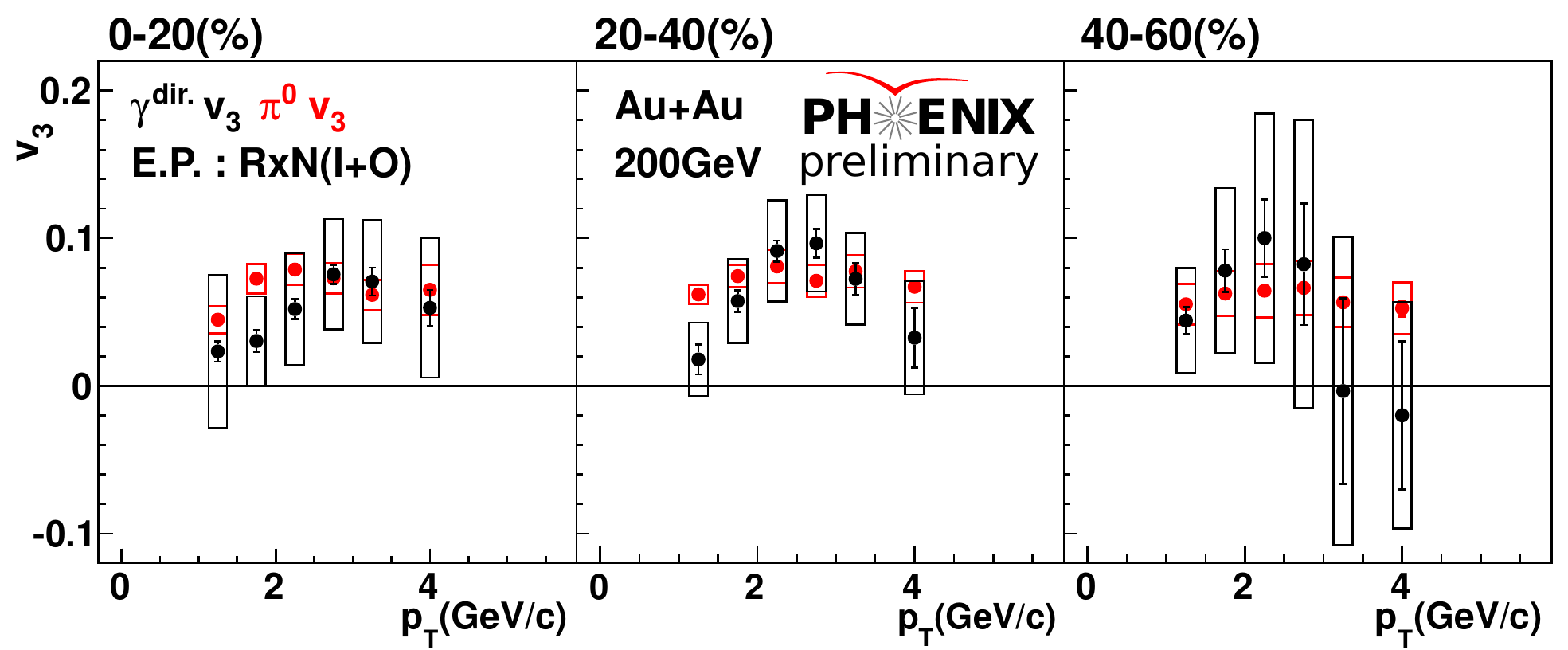}
\end{center}
\caption{(Color Online) The $v_3$ of direct photons and pions in Au+Au collisions at 200 GeV~\cite{Mizuno:14}. }
\label{fig:4}
\end{figure}

At this conference, STAR presented their first direct photon yields from dielectron measurements and found an excess in 0-80\% Au+Au over $p+p$ at $\sqrt{s_{_{NN}}} = 200$ GeV for $1\!<\!p_{T}\!<\!4$ GeV/$c$~\cite{Yang:14}. For $4\!<\!p_{T}\!<\!10$ GeV/$c$, the yields are consistent with the number of binary scaled $p+p$ measurements. The lack of the $\eta$ meson measurements at $p_{T}\!<\!2$ GeV/$c$ leads to large uncertainties in the hadronic cocktail components. This leads to large systematic uncertainties (greater than 50\%) in the direct virtual photon yields for $1\!<\!p_{T}\!<\!2$ GeV/$c$. Within these  uncertainties, the STAR measurements are consistent with the published PHENIX results~\cite{thermalphoton}, as shown in Fig.~\ref{fig:3}. 

Using external conversion technique, PHENIX measured the direct photon yields down to $p_{T}$ of 0.4 GeV/$c$. Excess yields in Au+Au over $p+p$ are observed at $\sqrt{s_{_{NN}}} = 200$ GeV, which are  exponential in $p_T$ with slope parameters independent of centrality~\cite{Mizuno:14}. It is found that the excess follows $N_{part}^{1.48\pm0.08\pm0.04}$ dependence. In addition, PHENIX observed that direct photons have a significant triangular flow $v_3$, similar to that of pions, as shown in Fig.~\ref{fig:4}.

In addition, ALICE reported that they were doing critical assessments for the measurements of $p_T$ spectrum and $v_2$ of direct photons in Pb+Pb collisions at 2.76 TeV. Conclusions should be drawn after assessments are completed~\cite{Bock:14}.

\subsection{The future improvements}
The excess $p_T$ spectrum of thermal photon in A+A collisions depends on the measurements of the direct photon spectra in A+A and $p+p$ collisions. The future improvements thus come from those from A+A and $p+p$. In particular, for A+A collisions, the measurement using internal conversion method would benefit from the precise $\eta$ measurement at $p_T\!<\!$ 2 GeV/$c$, while that using external conversion would benefit from more precise $\pi^{0,\pm}$ and $\eta$ measurements.  In $p+p$ collisions, the measurement of  direct photon at $p_T\!<\!$ 2 GeV/$c$ has large uncertainties at RHIC and does not exist at LHC. Future precise measurement is critical to get a solid physics picture of the excess $p_T$ yield.

In addition, it has been proposed that dilepton $v_2$
measurements
%will provide another independent way to study medium
%properties since dileptons provide two independent kinematic
%parameters: mass and $p_T$. Specifically, $v_2$
as a function of
$p_T$ in different mass regions would allow to probe the
properties of medium from a hadron-gas-dominated to a QGP-dominated scenario~\cite{Gale:07}.
At the previous and this Quark Matter, STAR reported the dielectron $v_2$
measurements from 200 GeV Au+Au collisions. Within uncertainties, the data are compatible with  known hadronic sources without QGP or hadron
gas thermal radiation~\cite{stardielectronv2}.  Much more data are needed for the dielectron $v_2$
measurement to provide additional sensitivity to study the thermal
radiation from the different phases in addition to the $\mu-e$ correlation measuring the charm correlation contribution. In the future, dimuon might provide an alternative approach considering the trigger capabilities~\cite{starmtdproposal}.

\section{Perspectives}
In QCD vacuum, the quark condensate has non-zero expectation value, which leads to spontaneous chiral symmetry breaking. This generates 99\% of the visible mass in the universe. In addition, the mass degeneracy between chiral partners, like $\rho(770)$ and $a_1(1260)$ mesons, is lifted. When the temperature and density is high enough, theoretical calculations predict that the chiral symmetry will be restored. The mass difference between  $\rho$ and $a_1$ will disappear. Experimentally, there are many efforts since decades to search for the signature of the chiral symmetry restoration. Since it is very difficult and challenging to measure $a_1$, the experimental efforts goes to the study of the modification of $\rho$ spectral function. An observed broadened and structureless spectral function would imply chiral symmetry restoration.

 \begin{figure}[htbp]
\begin{center}
\includegraphics[width=0.6\textwidth]{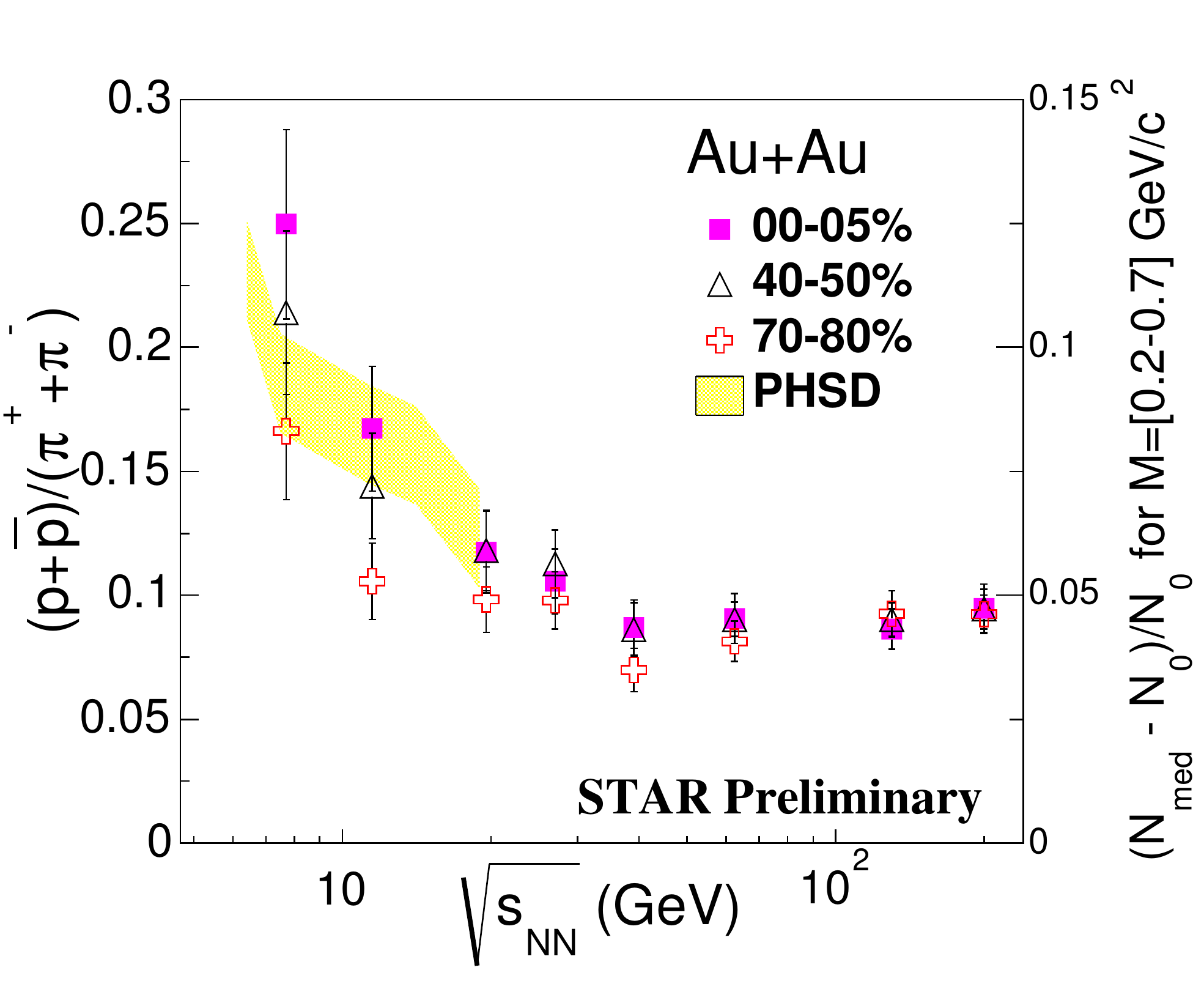}
\includegraphics[width=0.6\textwidth]{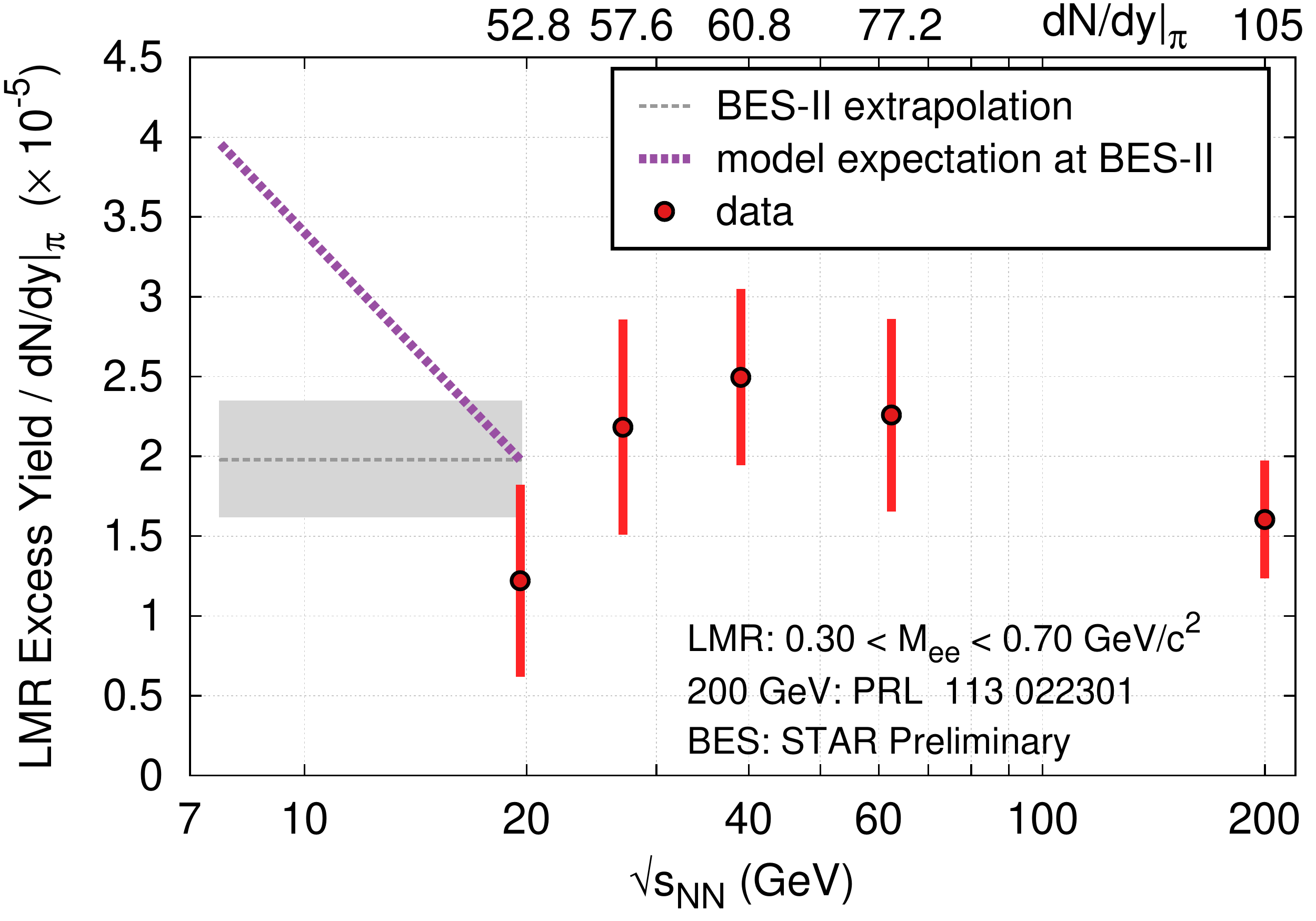}
\end{center}
\caption{(Color Online) (top panel) The $(p+\bar{p})/(\pi^{+}+\pi^{-})$ ratios as a function of energy in Au+Au collisions together with the excess yield predictions from a model~\cite{besII:14}. (bottom panel): The integral dielectron excess yields within STAR acceptance in the mass region 0.3-0.7 GeV/$c^2$ normalized by mid-rapidity pion yields in Au+Au collisions~\cite{Huck:14}. Data are measurements from 19.6 to 200 GeV. The rising line represents model expectations and the grey band is for the extrapolations assuming the normalized excess yields do not change down to 7.7 GeV. }
\label{fig:5}
\end{figure}

 It is found that a broadened $\rho$ spectral function~\cite{rapp:09}, describes
SPS dilepton data, consistently accounts for the STAR low mass
excess at 19.6, 27, 39, 62.4, and 200 GeV. Furthermore, it is found that coupling to baryons plays an essential role in the modification of $\rho$ spectral function in the hot, dense medium. 
The total baryon density, derived from the $(p+\bar{p})/(\pi^{+}+\pi^{-})$ ratios, does not change significantly from 17.3 to 200 GeV, as shown in the top panel of Fig.~\ref{fig:5}. Therefore, the measurements from 17.3 to 200 GeV probe the temperature and system evolution dependence. Figure~\ref{fig:5} (bottom panel) shows the integral dielectron excess yields within STAR acceptance in the mass region 0.3-0.7 GeV/$c^2$ normalized by mid-rapidity pion yields. No significant energy dependence of the normalized excess yields from 19.6 to 200 GeV is observed~\cite{Huck:14}.

In order to be sensitive to the total baryon density effect, one need to have measurements at lower energies. At 7.7 GeV, it is found that the total baryon density increases by a factor of two, as indicated in the top panel of Fig.~\ref{fig:5}.  Current data at 7.7 and 11.5 GeV at RHIC are not sufficient for dilepton analysis. The 
future measurements from the Beam Energy Scan Phase II at RHIC, will map out the dependence of modified $\rho$ spectral function on the total baryon density from 7.7 to 19.6 GeV Au+Au collisions~\cite{besII:14}.  The grey band shown in the bottom panel of Fig.~\ref{fig:5}  represents the precision projection if the normalized excess yield does not change. If it increases when the energy decreases as predicted by models, the uncertainties down to 7.7 GeV might be smaller. In addition, there exists 
a proposal of a new NA$60^{+}$ experiment at SPS to measure the excess dimuon mass spectra precisely at 6-20 GeV~\cite{Usai:14}. These, together with the future measurements from ALICE at LHC~\cite{Kohler:14}, HADES at SIS18~\cite{HADES:14}, and CBM at SIS100~\cite{FAIR:14} will provide a unique opportunity to study chiral symmetry restoration.

\section{Conclusions}
This edition of the Quark Matter conferences has seen a stream of results unprecedented in their quality and in their quantity, being presented by the experimental collaborations working at RHIC and at the LHC. Thermal photon and dilepton measurements from a broad beam energy range enable us to study the fundamental properties of QGP, of chiral symmetry restoration, and will provide stringent tests of the dynamical evolution scenarios  of relativistic nuclear collisions.\\

\noindent {\bf Acknowledgment} The work of of LR is supported in part by the U. S. Department of Energy under Contract No. DE-AC02-98CH10886 and under Early Career Research Program  Funding Award No. FWP$\#$2013-BNL-PO143.



%% References


%\bibliographystyle{elsarticle-num}



\end{document}